\documentclass[letterpaper, 10 pt, conference]{ieeeconf}  

\IEEEoverridecommandlockouts 
\usepackage{graphicx} 
\usepackage{amsmath}
\usepackage{amsfonts}
\usepackage{amssymb}
\usepackage{mathtools}
\usepackage{bm}
\usepackage{cite}
\usepackage{enumerate}

\newtheorem{thm}{Theorem}
\newtheorem{lemma}[thm]{Lemma}
\newtheorem{assumption}[thm]{Assumption}
\newtheorem{property}[thm]{Property}
\newtheorem{proposition}[thm]{Proposition}

\newtheorem{remark}[thm]{Remark}

\usepackage[colorlinks=true, linkcolor=blue]{hyperref}

\newcommand{\N}{\mathbb{N}}
\newcommand{\R}{\mathbb{R}}
\newcommand{\X}{\mathbb{X}}
\newcommand{\U}{\mathbb{U}}
\newcommand{\fK}{\mathcal{K}}
\newcommand{\Rp}{\mathbb{R}_{\geq 0}}
\newcommand{\Gt}{\tilde{\Gamma}}
\newcommand{\muNM}{\mu^\lambda_{N,M}}

\newcommand{\bu}{\boldsymbol{u}}
\newcommand{\intOmega}{\tilde{\Omega}}

\newcommand*{\QEDA}{\null\nobreak\hfill\ensuremath{\blacksquare}}

\makeatletter
\newcommand{\hiddentag}[2]{%
  \def\@currentlabel{#2}%
  \label{#1}%
}
\makeatother

\title{On the recursive feasibility and stability of MPC without terminal constraints}
\author{Mara Perez, Ignacio Sanchez, Filiberto Fele and Daniel Limon
	\thanks{Support from grants PID2022-142946NA-I00 and PID2022-141159OB-I00 funded by MICIU/AEI/ 10.13039/501100011033 and by ERDF/EU is gratefully acknowledged. F.~Fele also acknowledges support from
		grant RYC2021-033960-I funded by MICIU/AEI/ 10.13039/501100011033 and European Union NextGenerationEU/PRTR.}
        \thanks{Mara Perez is with the  Instituto de Desarrollo Tecnológico para la Industria Química (INTEC), CONICET - Universidad Nacional del Litoral (UNL), Santa Fe, Argentina. 
		{\tt\small mperez@fiq.unl.edu.ar}}
	\thanks{I.~Sanchez, F.~Fele and D.~Limon are with the Dept.~of Systems Engineering and Automation, 
		University of Seville, Av.~de los Descubrimientos s/n, 41092 Seville, Spain.
		{\tt\small \{isanchez10, ffele, dlm\}@us.es}}}

\begin{document}
	\maketitle

\begin{abstract}
In this paper, we analyze stability and recursive feasibility of a model predictive control (MPC) scheme that does not rely on terminal ingredients---such as terminal constraints or a terminal cost based on a terminal controller.
In the proposed formulation, we augment the cost function by a weighted finite tail which is optimized online, and we show it plays the role of a stabilizing terminal cost.
Under a relaxed, local dynamic programming condition on the tail, we certify a family of regions---parameterized by the tail's weight coefficient---on which the problem is recursively feasible and the origin is asymptotically stable.
These regions are larger as the weight increases: 
in particular, we show that any state that can be steered within the control horizon to a predefined neighborhood of the origin is contained in a certified region for a sufficiently large weight. 
Finally, we characterize the closed-loop performance over the certified region, bounding the suboptimality of the closed loop with respect to the infinite-horizon optimal cost. 
We conclude the discussion by illustrating these results on a simple numerical example.
\end{abstract}

\section{Introduction}

Model predictive control (MPC) is a well-established advanced control technique that provides stability guarantees, constraint satisfaction, and computational tractability for nonlinear systems~\cite{rawlings2020model,grune2013nonlinear}.
For closed-loop stability, the standard MPC formulation relies on the so-called terminal ingredients included in the cost function and the constraints. 
However, this requires the offline design of these ingredients~\cite{mayne2000survey}, which is often challenging, and simplifying assumptions are known to lead to overly conservative formulations (see discussion in~\cite{Mayne2013apol}).
Consequently, a significant research effort has been devoted to stabilizing MPC schemes that do not rely on such ingredients.

Different strategies have been proposed to circumvent the need for terminal ingredients.
Under standard assumptions, the authors of~\cite{limon2006stability} have shown that closed-loop stability can be achieved when the terminal constraint is removed. 
Specifically, they characterize a parameterized region in which the resulting optimal sequence implicitly satisfies the terminal constraint~\cite[Lemma~1]{limon2006stability} and certify invariance and stability for this region in the closed-loop using classic Lyapunov arguments.
This strategy still requires the design of the terminal ingredients, but the terminal constraint is not enforced.

Pursuing a more ambitious objective,~\cite{grune2012nmpc} considers an MPC framework without any terminal ingredient.
It relies on a relaxed version of the dynamic programming principle: instead of imposing a one-step decrease of the optimal value $V_N$ by an amount corresponding to exactly the stage cost (as in the infinite-horizon Bellman equation), only a decrease by a fraction $\alpha \in (0,1]$ of it is required.
A controllability-type bound on $V_N$ ensures $\alpha > 0$ for a sufficiently large horizon $N$, so that $V_N$ becomes a Lyapunov function, guaranteeing asymptotic stability.
Recursive feasibility, however, requires the state constraint set to be control invariant, which is very hard to ensure.
If invariance holds only locally around the equilibrium, recursive feasibility can still be derived from stability on sublevel sets of $V_N$ for long enough horizons; see also \cite[Section~8.3]{grune2013nonlinear}.
This local requirement is removed in~\cite{boccia2014stability}, at the expense of typically very large horizon bounds.

Stability with shorter horizons is achieved in~\cite{kohler2021stability}, where the control horizon is extended using a stabilizing control law defined within some neighborhood of the origin.
This formulation resembles the classical terminal-cost formulation and replaces the standard terminal penalty with the finite-horizon cost of a stabilizing control law, as proposed in~\cite{Magni2001}.
Although this perspective recovers the stability and feasibility properties of the classical MPC formulation, it still requires knowledge of a local stabilizing law on some set containing the origin. 

Stability without terminal constraints is obtained via a contraction argument in~\cite{alamir2018stability}. 
Shaping the cost by a penalty on the stage cost, which monotonically increases along the control horizon, is shown to be sufficient to stabilize the origin, provided that the penalty rate is sufficiently large. 

In this note, we develop an MPC formulation that does not require the design of explicit terminal ingredients.
Building on the scheme of~\cite{grune2012nmpc}, we split the finite-horizon cost into two sections and impose a suitable penalty weight on the tail.
We show how this tail plays the same role as a weighted terminal cost, making it possible to extend the sublevel-set arguments of~\cite{limon2006stability} to the proposed formulation. 
As a result, we obtain stability guarantees on an enlarged region while preserving recursive feasibility---without requiring the offline design of a terminal control law and a control Lyapunov function (CLF) associated with it.
Also, we show that the region of certified stability grows as the weight increases, covering the controllable set to a predefined neighborhood of the origin.
In addition, we determine suboptimality bounds for the closed-loop operation in relation to the infinite-horizon optimal cost.
We emphasize that, unlike the finite-tail formulation in~\cite{kohler2021stability}, our method does not require the availability of a stabilizing control law to generate the tail. 
Also, our design departs from the non-uniform weighting of~\cite{alamir2018stability}, as the penalty is concentrated on the tail rather than distributed along the horizon.

The remainder of the paper is organized as follows:
Section~II recalls the relaxed dynamic programming condition underlying MPC without terminal constraints, 
and presents the proposed formulation with a weighted finite tail.
Section~III characterizes certified regions of recursive feasibility and asymptotic stability, and derives closed-loop suboptimality bounds.
Section~IV illustrates the results on a numerical example, and Section~V concludes the paper.

\textit{Notation.}
A continuous function $\eta : \Rp \to \Rp$ is said to be of class $\mathcal{K}$ if it is strictly increasing and satisfies $\eta(0) = 0$.
If $\eta \in \mathcal{K}$ is also unbounded, $\eta$ is called a class $\mathcal{K}_\infty$-function.
Given a set $\mathbb{S}$ and $N\in\mathbb{N}$, $\mathbb{S}^N$ denotes the $N$-fold Cartesian product of $\mathbb{S}$, i.e., $\mathbb{S}^N = \underbrace{\mathbb{S}\times\cdots\times \mathbb{S}}_{N\text{ times}}$.

\section{Problem definition}

 \subsection{Setup}
Consider the following nonlinear discrete-time system:
\begin{equation}\label{eq:sys}
    x(k+1) = f(x(k),u(k)), \quad x(0) = x,
\end{equation}
where $x\in \R^n$ is the state, $u\in \R^m$ is the control input, $f:\R^n\times\R^m\to\R^n$ denotes the system dynamics, and $k\in\N_0$.
The state and input are constrained to a closed set $\X\subseteq\R^n$ and a compact set $\U\subseteq\R^m$, respectively. The origin is an equilibrium, i.e., $f(0,0)=0$ with $(0,0)\in \X\times\U$.

Given any state $x \in \X$ we say that the control sequence $\bu(x) \in \U^N$ is \emph{admissible} for~\eqref{eq:sys} if the resulting state trajectory is such that $x(k)\in\X$ for all $k = 0, \dots, N$, with $x(0) = x$.
Let $\U_N(x)$ denote the set of all admissible control sequences of length $N$ for a given initial state $x$, i.e., 
$$\U_N(x) \coloneqq \{\bu \in \U^N : x(k) \in \X, k = 0,\cdots,N, \text{ s.t.}~\eqref{eq:sys} \}.$$ 
Also, for any $\mathbb{S}\subseteq\X$ we define 
$$\X_N(\mathbb{S})\coloneqq\{x\in\X\colon \exists\,\bu\in\U_N(x)\text{ such that }x(N)\in\mathbb{S}\},$$ 
i.e., $\X_N(\mathbb{S})$ represents the $N$-step controllable set to $\mathbb{S}$.
In particular, $\X_N(\X)$ is the set of states having an admissible sequence of length $N$.

Based on an MPC formulation with no terminal constraints, our goal is to derive a state-feedback controller together with a certified region of states on which recursive feasibility and stability of the origin are guaranteed, that is, the closed-loop trajectory satisfies the constraints at all times and the state converges asymptotically to the origin.

Let us first consider a positive-definite stage-cost function $\ell\colon\X\times\U\to\R$, such that $\ell(0,0) = 0$, and define 
$$\ell^*(x) \coloneqq \min_{u \in \U_1(x)} \ell(x,u).$$ 

\begin{assumption}
\phantomsection
\label{ass:boundlstar}
    There exist functions $\xi_1,\xi_2\in\fK_\infty$ such that $\xi_1(\lVert x\rVert)\leq \ell^*(x)\leq\xi_2(\lVert x\rVert)$ for all $x\in\X$.
\end{assumption}

Assumption~\ref{ass:boundlstar} is standard for stabilizing controllers, commonly fulfilled by quadratic stage costs. 

\subsection{Stabilizing MPC without terminal constraints}
We next recall some sufficient conditions under which an MPC scheme without terminal constraints is stabilizing for a certain set $\Omega$ containing the origin.  The feedback law is recovered from the following optimization  problem (with the convention $V_M(x)=+\infty$ if $\U_M(x)=\emptyset$)
\begin{equation}\label{OCPN}
\begin{split}
    V_M(x) \coloneqq & \min_{\bu \in \U_M(x)}
    \sum_{k=0}^{M-1}\ell(x(k),u(k))
    \\
    \text{s.t. } & x(0) = x,\\
                  & x(k+1) = f(x(k),u(k)).
\end{split}
\end{equation}

In the following, we denote by $\bu^*_M(x)$ the optimal control sequence obtained as the solution of \eqref{OCPN}, and by $\mu_M(x)\coloneqq u^*_M(0)$ the resulting receding-horizon feedback policy.

\begin{assumption}
\phantomsection
\label{ass:VisaRCLF}
    For some $N_1\in\N$ and any $M\geq N_1$, there exist $\alpha \in (0,1]$ and $\beta>0$
    for which the optimal value function $V_M(x)$ satisfies
    \[ V_M(f(x,\mu_M(x))) - V_M(x) \leq - \alpha \ell(x,\mu_M(x)), \]
    for all $x\in \Omega\coloneqq\{x\in\X\colon V_M(x)\leq\beta\}$. 
    Moreover, $V_M$ is continuous at the origin.
\end{assumption}

Assumption~\ref{ass:VisaRCLF} provides the relaxed dynamic programming inequality as discussed in~\cite[Chapter~4]{grune2013nonlinear},
and implies that $\Omega$ is invariant for the closed loop $x(k+1) = f(x(k),\mu_M(x(k)))$, as $V_M(\cdot)$ is nonnegative and nonincreasing in time for all $x \in \Omega$.

The existence of $N_1$ and the computation of $\alpha$ from Assumption~\ref{ass:VisaRCLF} have been studied by Gr\"une and Pannek under certain controllability assumptions regarding the stage cost~\cite{grune2013nonlinear}.
In particular, if a stabilizing controller is assumed to be known on $\Omega$, $\alpha$ and $N_1$ can be computed explicitly from the bounds that this controller imposes~\cite{grune2012nmpc,kohler2021stability}. 

Overall, Assumption~\ref{ass:VisaRCLF} together with Assumption~\ref{ass:boundlstar} guarantee that the optimal cost function $V_M$ is a Lyapunov function (inside $\Omega$) which ensures the asymptotic stability of the origin.
This is stated next as an adaptation of~\cite[Proposition~3.4]{grune2012nmpc}.

\begin{proposition}
    Let Assumption~\ref{ass:boundlstar} and Assumption~\ref{ass:VisaRCLF} hold.
    Then $\mu_M(\cdot)$ renders the origin asymptotically stable. 
\end{proposition}

\subsection{MPC Formulation}

In this section, we study some basic properties that can be achieved by embedding a penalized finite-tail cost into an MPC formulation without terminal constraints.
In doing this, we show how the proposed controller relates to an MPC scheme with terminal cost $\lambda V_M(x(N))$, namely, the $\lambda$-weighted optimal cost of an $M$-horizon finite tail.

For $\lambda\ge1$, consider the following MPC optimization problem 
\begin{equation*}
\hiddentag{eq:PNM}{$P_{N,M}(x;\lambda)$}
\begin{split}
P_{N,M}(x;\lambda &):\\
     \min_{\bu \in \U_{N+M}(x)} &  \sum_{k=0}^{N-1}\ell(x(k),u(k))+\lambda\sum_{k=N}^{N+M-1}\ell(x(k),u(k))\\
    \text{s.t.}~&x(k+1) = f(x(k),u(k)),~x(0) = x,
\end{split}
\end{equation*}
where the control horizon is defined by $N\geq1$ and $M\ge N_1$, with $N_1$ as per Assumption~\ref{ass:VisaRCLF}.
Notice that the cost in~\ref{eq:PNM} can be equivalently interpreted as
\begin{equation*}
    J_{N}(x,\bu';\lambda)\coloneqq \sum_{k=0}^{N-1} \ell(x(k),u'(k)) + \lambda V_{M}(x(N)),
\end{equation*}
which is optimized over $\bu'\in\U_{N}(x)$. In other words, \ref{eq:PNM} can be understood as the MPC optimal control problem with horizon $N$ and 
 terminal cost $\lambda V_M(x(N))$---where $V_M(x(N))$ embeds the optimization problem defined in~\eqref{OCPN}.

Given $x\in\mathbb{X}$ and $\lambda\geq 1$, let $V_{N,M}(x;\lambda)$ denote the optimal value of \ref{eq:PNM}, attained for some optimal state and input trajectories $(\bm{x}^*,\bu^*)$ ($V_{N,M}(x;\lambda)=+\infty$ if no solution exists); finally, let $\muNM(x) \coloneqq {u}^*(0)$ be the corresponding receding-horizon feedback law.
Next, we study the closed-loop properties obtained from \ref{eq:PNM}, considering the relation between the penalizing factor $\lambda$ and the relaxation factor $\alpha$ from Assumption~\ref{ass:VisaRCLF}.\footnote{Unless otherwise indicated, throughout the paper $\alpha$ refers to the constant associated with \eqref{OCPN} with horizon $M$. A similar convention will be used for $\beta$ and $\Omega$.}
Of particular interest is the case $\lambda \geq 1/\alpha$, which makes the weighted finite-tail cost $\lambda V_M(\cdot)$ a CLF over $\Omega$ under Assumption~\ref{ass:VisaRCLF} which fulfills the stabilizing conditions presented in~\cite[Assum.~2.14]{rawlings2020model}. 

\begin{property}
\phantomsection
\label{prop:VMisCLF}
    Let Assumption~\ref{ass:VisaRCLF} hold, and select $M\geq N_1$. 
    In addition, consider any $\lambda \geq 1/\alpha$. 
    Then $\lambda V_M(x)$ satisfies 
    \begin{equation}\label{eq:VM-CLF}
         \lambda V_{M}(f(x,\mu_M(x))) - \lambda  V_{M}(x) \leq - \ell(x,\mu_M(x)),
    \end{equation}
    for all $x\in\Omega$.
\end{property}

\begin{proof}
    By Assumption~\ref{ass:VisaRCLF}, $V_M$ satisfies
    \begin{equation*}
        V_{M}(f(x,\mu_M(x))) - V_{M}(x) \leq -\alpha \ell(x,\mu_M(x)),
    \end{equation*} 
    provided that $x \in\Omega$.
    Since $\lambda \geq 1/\alpha$, we have that
    \begin{multline*}
        \lambda(V_{M}(f(x,\mu_M(x))) - V_{M}(x)) \\
        \leq -\lambda\alpha \ell(x,\mu_M(x)) \leq -\ell(x,\mu_M(x)),
    \end{multline*}
    which implies that~\eqref{eq:VM-CLF} is satisfied for all $\lambda\geq 1/\alpha$.
\end{proof}

Compared to the standard MPC scheme with terminal ingredients~\cite{mayne2000survey}, the proposed formulation drops the terminal set constraint---no constraint enforces $x^*(N)\in\Omega$ in \ref{eq:PNM}---and the terminal cost is replaced by a penalized finite tail $\lambda V_M(\cdot)$, which is a CLF only locally on $\Omega$.

Now we can reformulate the goals of this work: we aim to provide conditions under which \ref{eq:PNM} is recursively feasible and the origin is asymptotically stable in the closed loop.
The main difficulty is that, without a terminal constraint, nothing in~\ref{eq:PNM} forces the predicted terminal state into $\Omega$. 
Yet $\lambda V_M(\cdot)$ is guaranteed to be a CLF by Assumption~\ref{ass:VisaRCLF} only in $\Omega$. 
If $x^*(N)\in\Omega$ and $\lambda\geq 1/\alpha$, then $\lambda V_M(\cdot)$ acts as a standard terminal cost (Property~\ref{prop:VMisCLF}), and recursive feasibility and asymptotic stability can follow from standard Lyapunov arguments for MPC with terminal ingredients~\cite{rawlings2020model}. 
Hence, the key step is to guarantee that $x^*(N)\in\Omega$ without enforcing it explicitly through a constraint.

To this end, we adopt the approach of~\cite{limon2006stability}, which was developed for MPC with a CLF terminal cost and no terminal constraint. It relies on two facts. First, if the optimal predicted trajectory does not end in $\Omega$, then it never visits $\Omega$~\cite[Lemma~1]{limon2006stability}.
Second, since the stage cost is bounded from below by some $d>0$ outside $\Omega$, such a trajectory incurs a large cost. 
As a result, the terminal constraint is implicitly satisfied on a sublevel set of the optimal cost, and this set can be enlarged by increasing the weight of the terminal cost~\cite[Theorems~2 and~3]{limon2006stability}.

Following these observations and Property~\ref{prop:VMisCLF}, we focus on the set of admissible initial states $x\in\X_N(\X)$ from which $\Omega$ is reached in no more than $N$ steps by the solutions of~\ref{eq:PNM}.
By definition of $\Omega$ in Assumption~\ref{ass:VisaRCLF}, this set can be equivalently defined in terms of $V_M$:
\begin{equation}\label{eq:Gamma}
\Gamma_{N}(\lambda) \coloneqq \{x\in\X_N(\X)\colon V_{M}(x^*(N))\leq\beta\},    
\end{equation} 
where we emphasize the role of $\lambda$ as a tuning parameter for \ref{eq:PNM}.

The following lemma provides sufficient conditions under which the control law $\muNM(\cdot)$ affords standard closed-loop stability properties over $\Gamma_{N}(\lambda)$.

\begin{lemma}
\phantomsection
\label{lem:F>VN}
    Let Assumption~\ref{ass:VisaRCLF} hold and consider $P_{N,M}(x;\lambda)$
    with $M\geq N_1$ and $\lambda \geq 1/\alpha$.
    Then
    \begin{enumerate}[i)]
        \item $V_{N+1,M}(x;\lambda) \leq V_{N,M}(x;\lambda)$,
        $\forall x \in \Gamma_{N}(\lambda) $, $N\in\N_0$;        
        \item $V_{N,M}(x;\lambda) \leq \lambda V_M(x)$, $\forall x\in\Omega$, $N\in\N_0$.
    \end{enumerate}
\end{lemma}

\begin{proof}
    The proof follows from classical results since Property~\ref{prop:VMisCLF} implies $\lambda V_{M}(x)$ is a CLF on $\Omega$ for any $\lambda \geq 1/\alpha$; see, e.g., \cite[Proposition~2.8]{rawlings2020model} or~\cite[Lemma~5.12]{grune2013nonlinear}. For completeness, the essential idea underlying the proof is outlined in the Appendix.
\end{proof}

Property~\ref{prop:VMisCLF} allows the finite-tail cost to be interpreted as a terminal cost that possesses the properties of a CLF. 
Furthermore, Lemma~\ref{lem:F>VN} shows that \ref{eq:PNM} recovers the standard upper bound on the finite-horizon optimal value.
It follows that the proposed formulation inherits the stability and performance properties typically associated with MPC schemes incorporating terminal costs.

In the next section, we deepen our analysis by identifying certified stability regions and establishing performance bounds for the closed loop.

\section{Theoretical analysis}

\subsection{Certified Stability Regions}
Next, we characterize the inner approximation $\Gt_N^d(\lambda) \subseteq \Gamma_{N}(\lambda)$ and explore the role of the parameter $\lambda$ in its definition.
Then, we will certify recursive feasibility and stability for the proposed controller on the region $\Gt_N^d(\lambda)$. 
It is known that a larger domain of attraction can be obtained by increasing the control horizon $N$~\cite{rawlings2020model,kohler2021stability}, albeit this comes at the expense of computational resources.
In the last part of this subsection, we focus instead on the effect that the parameter $\lambda$ has on the size of the certified domain of attraction.

We begin by formalizing a property of the optimal trajectory from $P_{N,M}(x;\lambda)$ with respect to the set $\Omega$ defined in Assumption~\ref{ass:VisaRCLF}.

\begin{lemma}
\phantomsection
\label{lem:xNinOmega}
    Let Assumption~\ref{ass:VisaRCLF} hold, and consider the optimization problem \ref{eq:PNM} with $M\geq N_1$, $\lambda\geq 1/\alpha$, and $x\in \X_N(\X)$.
    If $(\bm{x}^*,\bu^*)$ is an optimal solution and $x^*(N)\notin\Omega$, then $x^*(j)\notin\Omega$ for all $j = 0,\dots,N-1$.
\end{lemma}

\begin{proof}
    Following the proof of~\cite[Lemma~1]{limon2006stability}, we proceed by contradiction.
    For this, suppose that there exists $j$, with $0\leq j\leq N-1$, such that $x^*(j)\in\Omega$.
    By the dynamic programming principle, the tail $\{u^*(j),\dots,u^*(N-1)\}$ is optimal for $P_{N-j,M}(x^*(j);\lambda)$, hence
    \[ V_{N-j,M}(x^*(j);\lambda) = \sum_{k=j}^{N-1}\ell(x^*(k),u^*(k)) + \lambda V_M(x^*(N)). \]
    Since $x^*(j)\in\Omega$, Lemma~\ref{lem:F>VN}-ii applies and, as $\ell\geq0$, we obtain
    \begin{align*}
        \lambda V_M(x^*(j)) &\geq V_{N-j,M}(x^*(j);\lambda)\\
        &\geq \lambda V_M(x^*(N))\\
        &> \lambda\beta,
    \end{align*}
    where the last inequality follows from $x^*(N)\notin\Omega$, i.e., $V_M(x^*(N))>\beta$.
    This implies $V_M(x^*(j))>\beta$, hence $x^*(j)\notin\Omega$, which contradicts the hypothesis.
\end{proof}

We can now present our main stability result. Specifically, we characterize computable inner approximations of $\Gamma_{N}(\lambda)$ over which \ref{eq:PNM} is recursively feasible in closed loop and the origin is an asymptotically stable equilibrium.

\begin{thm}
\phantomsection
\label{thm:MPC-stability}
    Let Assumptions~\ref{ass:boundlstar} and \ref{ass:VisaRCLF} hold and select $N\geq 1$, $M \geq N_1$ and $\lambda\geq1/\alpha$.
    Let $d>0$ be such that $\ell(x,u)\geq d$ for all $x\notin\Omega$ and all $u\in\U_1(x)$.     
    Then, the receding-horizon feedback $\muNM(\cdot)$ obtained from $P_{N,M}(\cdot;\lambda)$ asymptotically stabilizes the origin over the region defined by 
        \begin{equation}\label{eq:Gt}
        \Gt_{N}^d (\lambda) \coloneqq \left\{x\in\X_N(\X)\colon V_{N,M}(x;\lambda) \leq Nd + \lambda\beta\right\}.
        \end{equation}
\end{thm}
\vspace{6pt}
\begin{proof}
    The proof is developed in three parts, which show that i) $\Gt_N^d(\lambda)$ is contained in $\Gamma_{N}(\lambda)$, ii) $\Gt_N^d(\lambda)$ is invariant and iii) the optimal cost $V_{N,M}(x;\lambda)$ is a Lyapunov function satisfying $V_{N,M}(f(x,\muNM(x));\lambda) \leq V_{N,M}(x;\lambda) - \ell(x,\muNM(x))$ for all $x \in \Gt_N^d(\lambda)$.
    
    i) To show the inclusion of $\Gt_N^d(\lambda)$ in $\Gamma_{N}(\lambda)$, we demonstrate that for any $x\in\Gt_{N}^d(\lambda)$, the solution of $P_{N,M}(x;\lambda)$ satisfies $x^*(N)\in\Omega$. First, note that the set $\Gt_{N}^d(\lambda)$ is well defined: a value of $d$ fulfilling the hypothesis exists by assumption, since $\Omega$ contains a neighborhood of the origin. 
    We then proceed by contradiction, and suppose that $x^*(N)\notin\Omega$. 
    Thus, we have $\lambda V_M(x^*(N))>\lambda\beta$ and---by Lemma~\ref{lem:xNinOmega}---$x^*(j)\notin\Omega$ for $j = 0,\dots,N-1$, which by definition of $d$ implies $\ell(x^*(j),u^*(j))\geq d$ for all $j = 0,\dots,N-1$. It follows that
    \begin{align*}
        V_{N,M}(x;\lambda) &= \sum_{k=0}^{N-1}\ell(x^*(k),u^*(k)) + \lambda V_M(x^*(N))\\
        &> Nd + \lambda\beta,
    \end{align*}
    which contradicts $x\in\Gt_{N}^d(\lambda)$.
    Therefore $x^*(N)\in\Omega$, hence $\Gt_N^d(\lambda)\subseteq \Gamma_{N}(\lambda)$.

    ii) We prove that $\Gt_N^d(\lambda)$  is a positively invariant set for the closed loop, i.e., $x\in\Gt_N^d(\lambda)$ implies $f(x,\muNM(x))\in\Gt_N^d(\lambda)$.
    From part i), we know $x\in\Gt_N^d(\lambda)\subseteq \Gamma_{N}(\lambda)$, therefore  $x^*(N)\in\Omega$.
    Note that the sequence $\{x^*(1),\dots,x^*(N)\}$ is still optimal for $P_{N-1,M}(x^*(1);\lambda)$: we thus have $x^*(1) = f(x,\muNM(x)) \in \Gamma_{N-1}(\lambda)$.
    Hence, we can apply Lemma~\ref{lem:F>VN}, yielding 
    \begin{align}
        V_{N,M}(f(x,\muNM(x));\lambda) &\leq V_{N-1,M}(f(x,\muNM(x));\lambda)\nonumber\\
                                   & = V_{N,M}(x;\lambda) - \ell(x,\muNM(x))\nonumber\\
                                   &\leq Nd+\lambda\beta,\label{eq:thm1_proof_bound1}
    \end{align}
    for all $N\geq 1$, where the last inequality follows from the positive definiteness of $\ell$ and the fact that $x\in\Gt_N^d(\lambda)$. 
    By~\eqref{eq:Gt}, this implies $f(x,\muNM(x))\in\Gt_N^d(\lambda)$.
    
    iii) We show that $V_{N,M}(x;\lambda)$ is a Lyapunov function over $\Gt_N^d(\lambda)$. 
    For this, we first notice from \eqref{eq:thm1_proof_bound1} that the difference $V_{N,M}(f(x,\muNM(x));\lambda) - V_{N,M}(x;\lambda)$ is strictly negative along any closed-loop trajectory starting from any $x\in\Gt_N^d(\lambda)$, $x\neq 0$. Furthermore, by Assumption~\ref{ass:boundlstar}, there exists $\xi_1\in\fK_\infty$ which bounds $V_{N,M}(x;\lambda)$ from below, i.e.,
    \[ \xi_1(\lVert x\rVert) \leq \ell^*(x)  \leq \ell(x,\muNM(x)) \leq V_{N,M}(x;\lambda), \]
    that holds for all $x\in\Gt_N^d(\lambda)$. Finally, by Lemma~\ref{lem:F>VN}, $V_{N,M}(\cdot; \lambda)$ is upper bounded by $\lambda V_M(x)$, which is itself a Lyapunov function in $\Omega\subseteq\Gt_N^d(\lambda)$. 
    With $V_{N,M}(0;\lambda)=0$, this implies that $V_{N,M}(x;\lambda)$ is a Lyapunov function for the closed loop on the positively invariant set $\Gt_N^d(\lambda)$, thus guaranteeing asymptotic stability of the origin with a certified region $\Gt_N^d(\lambda)$.    
\end{proof}

Now we are ready to show two important features of the proposed scheme:
i) the certified regions $\Gt_N^d(\lambda)$ are nested with respect to $\lambda$ and ii), as $\lambda$ grows, they increasingly cover the set of states that can be steered within $N$ steps to
 \[
 \intOmega \coloneqq \{x\in\X\colon V_M(x)<\beta\}\subseteq \mathrm{int}\,\Omega.
 \]
 Consequently, starting from any $x\in\X_N(\intOmega)\subseteq \X_N(\Omega)$, the receding-horizon law $\muNM(\cdot)$ asymptotically stabilizes the origin provided that $\lambda$ is chosen large enough.
These facts are formalized next.

\begin{lemma}
\phantomsection
\label{lem:pointwise}
    Let Assumptions~\ref{ass:boundlstar} and~\ref{ass:VisaRCLF} hold.
    Then, for any $x\in\X_N(\intOmega)$, there exists a finite $\lambda\geq1/\alpha$ such that $x\in\Gt_N^d(\lambda)$.
\end{lemma}

\begin{proof}
    The proof adapts the argument of~\cite[Theorem~3]{limon2006stability}.
    Let $x\in\X_N(\intOmega)$. 
    Then there exists $\bu\in\U_N(x)$ whose associated trajectory satisfies $x(N)\in\intOmega$, that is, $V_M(x(N))<\beta$. 
    Since $V_M(x(N))<\infty$, we have $\U_M(x(N))\neq\emptyset$.
    Concatenating $\bu$ with $\bu_M^*(x(N))$ therefore yields an admissible sequence in $\U_{N+M}(x)$. 
    Using this sequence together with the corresponding state trajectory resulting from \eqref{eq:sys}, and letting $J(\bu)\coloneqq\sum_{k=0}^{N-1}\ell(x(k),u(k))$, the cost $J(\bu)+\lambda V_M(x(N))$ is well defined, and lower-bounded by $V_{N,M}(x;\lambda)$ due to optimality of the latter.
    Moreover, $J(\bu)+\lambda V_M(x(N)) \leq Nd+\lambda\beta$,
    holds for any $\lambda$ such that
    \[
    \lambda\geq\max\left\{\frac{1}{\alpha},\ \frac{J(\bu)-Nd}{\beta-V_M(x(N))}\right\},
    \]
    after noticing that $\beta-V_M(x(N))>0$.
    Thus, $V_{N,M}(x;\lambda)\leq Nd+\lambda\beta$, and with $d$ defined as in Theorem~\ref{thm:MPC-stability}, this is equivalent to $x\in\Gt_N^d(\lambda)$, concluding the proof.
\end{proof}

Lemma~\ref{lem:pointwise} shows that every state in $\X_N(\intOmega)$ lies in a certified region $\Gt_N^d(\lambda)$ for a sufficiently large, finite value of $\lambda$. 
With the following theorem, we demonstrate that the certified regions are nested, i.e., non-decreasing with respect to inclusion as $\lambda$ increases. Combined with Lemma~\ref{lem:pointwise}, this characterizes their limit as $\lambda\to\infty$.

\begin{thm}
\phantomsection
\label{thm:monotonicityGamma}
    Let Assumptions~\ref{ass:boundlstar} and~\ref{ass:VisaRCLF} hold, and consider~\ref{eq:PNM} with $N\geq1$ and $M\geq N_1$.
    Then, i) $\Gt_N^d(\lambda_1) \subseteq \Gt_N^d(\lambda_2)$ for all $\lambda_2\geq \lambda_1 \geq 1/\alpha$.
    Moreover, ii) the certified regions satisfy
    \[
    \X_N(\intOmega)\subseteq \bigcup_{\lambda\ge 1/\alpha}\Gt^d_N(\lambda)
        =\lim_{\lambda\to\infty}\Gt^d_N(\lambda)\subseteq \X_N(\Omega).
    \]
\end{thm}

\begin{proof}
    For i), we follow~\cite[Theorem~2]{limon2006stability}. 
    Let $x\in\Gt_N^d(\lambda_1)$ and consider the solution $(\bm{x}^*,\bm{u}^*)$ from $P_{N,M}(x;\lambda_1)$.
    Then,
    \begin{align*}
        V_{N,M}(x;\lambda_1) &= \sum_{k=0}^{N-1}\ell(x^*(k),u^*(k)) + \lambda_1 V_M(x^*(N))\\
                             &= \sum_{k=0}^{N-1}\ell(x^*(k),u^*(k))+ \lambda_2 V_M(x^*(N)) \\[-0.7em]
                             & \hspace{3cm} - (\lambda_2 - \lambda_1) V_M(x^*(N))\\
                             &\geq V_{N,M}(x;\lambda_2) - (\lambda_2 - \lambda_1) V_M(x^*(N)),
    \end{align*}
    where the last inequality is due to $(\bm{x}^*,\bm{u}^*)$ being optimal for $P_{N,M}(x;\lambda_1)$ but not necessarily for $P_{N,M}(x;\lambda_2)$.
    Since $x\in\Gt_N^d(\lambda_1)\subseteq \Gamma_N(\lambda_1)$, we have $x^*(N)\in\Omega$ and $V_M(x^*(N))\leq\beta$.
    Hence, for $\lambda_2 \geq \lambda_1$, 
    \begin{align*}
        V_{N,M}(x;\lambda_2) &\leq V_{N,M}(x;\lambda_1) + (\lambda_2 - \lambda_1) \beta\\
                             &\leq Nd + \lambda_1\beta + (\lambda_2 - \lambda_1) \beta\\
                             &= Nd + \lambda_2\beta,
    \end{align*}
    which implies $x\in\Gt_N^d(\lambda_2)$.

    For part ii), first note that the family $\{\Gt_N^d(\lambda)\}_{\lambda\geq1/\alpha}$ is non-decreasing with respect to inclusion, by i). 
    Hence the union equals the limit of $\Gt_N^d(\lambda)$ as $\lambda\to\infty$.
    The first inclusion follows from Lemma~\ref{lem:pointwise}.
    From Theorem~\ref{thm:MPC-stability}-i), $\Gt_N^d(\lambda) \subseteq \Gamma_N(\lambda)$ and $\Gamma_N(\lambda) \subseteq \X_N(\Omega)$ by definition, proving ii).
    \end{proof}

We note that under certain assumptions, e.g., continuity and strict quasiconvexity of the optimal cost $V_M$, the above results become tighter as $\intOmega$ coincides with $\mathrm{int}\,\Omega$.

\subsection{Suboptimality bounds}

Let $J_{\infty}^{cl}(x,\muNM;\lambda)$ denote the closed-loop cost incurred by applying $P_{N,M}(x;\lambda)$ in a receding-horizon fashion.
This satisfies the following bounds. 

\begin{proposition}\label{prop:MPCperformance}
    Let Assumptions~\ref{ass:boundlstar} and ~\ref{ass:VisaRCLF} hold and consider $\lambda\ge 1/\alpha$.
    Then, the MPC feedback obtained from \ref{eq:PNM} 
    is such that
    \begin{multline}\label{eq:prop_comparison}
        V_\infty(x)\le J_{\infty}^{cl}(x,\muNM;\lambda) \leq V_{N,M}(x;\lambda) \\ 
        \leq \lambda V_{N,M}(x;1)
        \leq\lambda V_\infty(x),
    \end{multline}
    for all $x\in\Gt_N^d(\lambda)$, where $V_\infty(x) = \lim_{M\to\infty} V_M(x)$.
\end{proposition}

\begin{proof}
    Let $\phi(x,j)$ denote the state of the closed-loop trajectory obtained at time $j$ by applying $P_{N,M}(\cdot;\lambda)$ in receding horizon starting from $\phi(x,0)=x$ (note that this is not necessarily the same as the \emph{predicted} trajectory denoted earlier by $x^*(j)$), and let $\ell(\phi(x,j),u(j))$ be the corresponding stage cost.
    The first inequality in \eqref{eq:prop_comparison} follows by definition of $V_\infty$, and the second one from 
    Theorem~\ref{thm:MPC-stability}, which implies that $V_{N,M}(x;\lambda)$ is a Lyapunov function satisfying
    \begin{multline*}
        V_{N,M}(\phi(x,j);\lambda) - V_{N,M}(\phi(x,j+1);\lambda) \\
            \geq  \ell(\phi(x,j),\muNM(\phi(x,j))),    
    \end{multline*}
    for all $j\in\N$ and all $x\in\Gt_N^d(\lambda)$.
    Summing the above inequality over $j\in\mathbb{N}$ yields 
    \begin{multline*}
       V_{N,M}(x;\lambda) \ge \sum_{j=0}^\infty  \ell(\phi(x,j),\muNM(\phi(x,j))) \\ 
        \eqqcolon J_{\infty}^{cl}(x,\muNM;\lambda),
    \end{multline*}
    since $\lim_{j\to\infty} V_{N,M}(\phi(x,j);\lambda) = 0$, which holds by Theorem~\ref{thm:MPC-stability}.
    The third inequality in \eqref{eq:prop_comparison} follows from the dynamic programming principle, i.e.:
    \begin{align*}
        V_{N,M}(x;\lambda) &= \min_{\bu\in\U_N(x)}\sum_{k=0}^{N-1}\ell(x(k),u(k)) + \lambda V_M(x(N))\\
        & \le \min_{\bu\in\U_N(x)}\lambda\bigg(\sum_{k=0}^{N-1}\ell(x(k),u(k)) + V_M(x(N))\bigg)\\
        & = \lambda V_{N,M}(x;1),
    \end{align*}
    which holds since $\lambda\geq 1$ by Assumption~\ref{ass:VisaRCLF}.
    The last inequality in \eqref{eq:prop_comparison} follows by definition. 
\end{proof}

\begin{remark}
The third term in \eqref{eq:prop_comparison} recovers the standard closed-loop cost bound for MPC based on terminal ingredients; by Lemma~\ref{lem:F>VN}, a tighter bound can be achieved by increasing $N$.
On the other hand, for fixed $N$ and $M$, the size of $\Gt_N^d(\lambda)$ can be increased through $\lambda$. In particular, our approach does not require Assumption~\ref{ass:VisaRCLF} to hold over the entire domain of attraction of the controller: instead, our results allow $\Omega$ to be an arbitrarily small sublevel set of $V_M$.
\end{remark}

\section{Numerical example}
We illustrate the applicability of the proposed scheme on the linear system $x(k+1) = A x(k) + B u (k)$, with $A = [1,1;0,2]$, $B = [1,0;0,1]$; state and input constraints are given by $|x_1| \leq 10, |x_2| \leq 1$, $|u_1|+|u_2| \leq 1$.
A quadratic stage cost is considered, $\ell(x,u) = x^\top Q x + u^\top R u$, with $Q = [100,0;0,1]$ and $R = [1,0;0,100]$.
Computing the finite-horizon LQR through the Riccati recursion $P_0=0$, $P_j = Q + A^\top P_{j-1}A - A^\top P_{j-1}B(R+B^\top P_{j-1}B)^{-1}B^\top P_{j-1}A$ for increasing horizons $j$, we find that the resulting closed loop $A+BK_j$ is Schur from $j=6$ onwards.
We therefore take a finite tail of length $M=6$, with $V_M(x)=x^\top P_6 x$ and $\Omega=\{x\colon x^\top P_6 x\leq\beta\}$---where $\beta$ is the largest value for which no constraint is active along the $M$-step trajectory of the 6-horizon LQR---and choose $N=10$.
This way, the proposed scheme certifies the stability regions with a combined horizon of $N+M=16$.
By numerical evaluation, we obtain $d = 0.0111$, $\alpha = 0.7608$ and $\beta = 2.8828$.

\begin{figure}
    \centering
    \includegraphics[width=1\columnwidth]{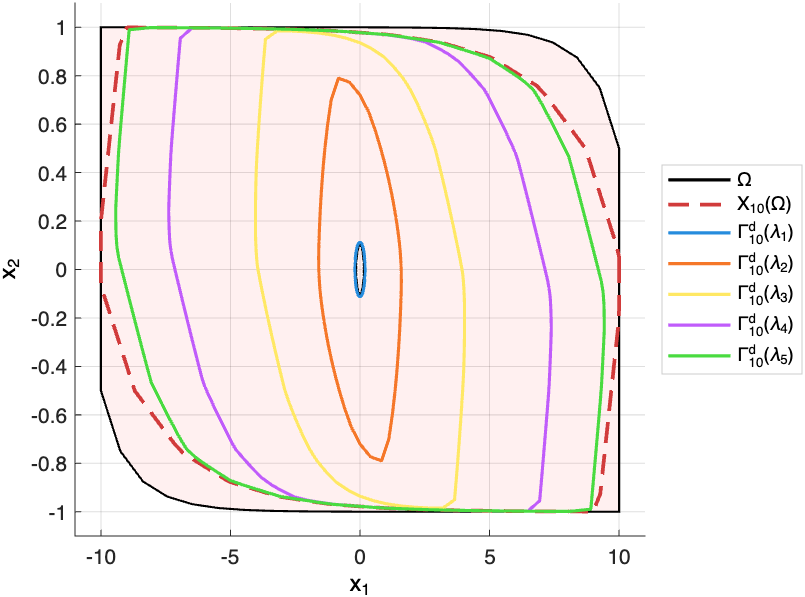}
    \caption{Sets of interest for the example with $M = 6$ and $N = 10$: Maximal control invariant set (shaded), $\Omega$ (black), 10-step controllable set to $\Omega$ (dashed), and certified domains of attraction $\Gt_N^d(\lambda_j)$ with $d = 0.0111$, and increasing values of $\lambda_j$,  $(\lambda_1,\dots,\lambda_5) = (1/\alpha,100, 1000, 5000,10000)$, where $1/\alpha = 1.3144$; the certified regions grow with $\lambda$.}
    \label{fig:Sets}
\end{figure}

Figure~\ref{fig:Sets} shows a selection of significant sets for this system: the maximal invariant set, $\Omega$, $X_N(\Omega)$ and $\Gt_N^d(\lambda)$ for selected values of $\lambda$.
These sets are consistent with the presented theoretical results.
First, we verify that the certified regions are nested, i.e., $\Gt_N^d(\lambda_1)\subseteq\Gt_N^d(\lambda_2)$ for $\lambda_1\leq\lambda_2$: moreover, $\Gt_N^d(\lambda)$ grows as $\lambda$ increases. 
Second, we verify that $\Gt_N^d(\lambda)$ is contained in $\X_N(\Omega)$, as expected, since $\X_N(\Omega)$ is the domain of attraction of the MPC controller with terminal constraint $x(N)\in\Omega$ and same horizon $N$.

\section{Conclusions}
In this work, we have presented an MPC scheme without terminal ingredients and produced certified regions of stability.
These regions are larger as the penalization on the finite tail increases, even for a fixed control horizon $N$.
We have also given closed-loop performance bounds that depend only on the local stability properties with respect to the finite tail. 
A numerical example illustrates some certified stability regions obtained for very short horizons using these techniques.

\appendix
\subsection{Proof of Lemma~\ref{lem:F>VN} (sketch)}
\label{lem:F>VN_proof}
    Let the pair $(\bm{x}^*,\bu^*)$ denote a solution of $P_{N,M}(x;\lambda)$ resulting in $V_{N,M}(x;\lambda)$.
    By the hypothesis of part i), $x^*(0) = x \in \Gamma_{N}(\lambda)$, hence $x^*(N) \in \Omega$. 
    By the dynamic programming principle, we have $u^*(N) =\mu_M(x^*(N))$ and, since $\Omega$ is positively invariant under $\mu_M$, $x^*(N+1) = f(x^*(N),\mu_M(x^*(N))) \in \Omega$, and therefore $V_M(x^*(N+1))<\infty$.
    Let $\hat{\bm{u}}\in\U_M(x^*(N+1))$ denote an optimal input sequence for $V_M(x^*(N+1))$.
    Then, a feasible solution for $P_{N+1,M}(x;\lambda)$ can be constructed using the input
    \begin{displaymath}
        \tilde{\bm{u}} = \{u^*(0),\dots,u^*(N),\hat{u}(0),\dots,\hat{u}({M-1})\},
    \end{displaymath}
    with (suboptimal) cost ${J}_{N+1}(x,\tilde{\bm{u}};\lambda)$ that fulfills
    \begin{displaymath}
    \begin{split}
        {J}_{N+1}(x,\tilde{\bm{u}};\lambda) &= \!\!\! \sum_{k=0}^{(N+1)-1} \!\!\! \ell(x^*(k),u^*(k)) + \lambda V_M(x^*(N+1))\\
        &= \sum_{k=0}^{N-1} \ell(x^*(k),u^*(k)) + \ell(x^*(N),u^*(N)) \\
        & \qquad \qquad + \lambda V_M(x^*(N+1))\\
        & \leq \sum_{k=0}^{N-1} \ell(x^*(k),u^*(k)) 
        + \lambda V_M(x^*(N))\\
        & = V_{N,M}(x;\lambda),
        \end{split}
    \end{displaymath}
    where the inequality follows from $\lambda V_M(x)$ being a CLF in $\Omega$.
    Therefore, since $ {V}_{N+1,M}(x;\lambda) \leq  {J}_{N+1}(x,\tilde{\bm{u}};\lambda)$, we have
    ${V}_{N+1,M}(x;\lambda) \leq {V}_{N,M}(x;\lambda)$, which proves part i).

    For part ii), let $x\in\Omega$ and apply the feedback $\mu_M$ for $N$ steps starting from $x$. By positive invariance of $\Omega\subseteq\X$ the resulting sequence is admissible and $x(N)\in\Omega$, and by Property~\ref{prop:VMisCLF}, $\sum_{k=0}^{N-1}\ell(x(k),\mu_M(x(k)))+\lambda V_M(x(N))\leq\lambda V_M(x)$. Hence $V_{N,M}(x;\lambda)\leq\lambda V_M(x)$. \QEDA

\bibliography{Bib} 
	
\end{document}